# A Novel RF Energy Harvesting Module Integrated on a Single Substrate

Monika Mathur[1], Ankit Agarawal[2], Ghanshyam Singh[3], S. K. Bhatnagar[4]

**Department of Electronics and Communication**
[1,2,4] Swami Keshvanand Institute of Technology , Management and Gramothan, Jaipur, Rajasthan, INDIA.
[1,3] Malaviya National Institute of Technology Jaipur, Rajasthan, INDIA
Email id: [1]monikamathur16@ gmail.com, [2]ankitsagarwal@ gmail.com, [3]gschoudhary75@ gmail.com,
[4]bhatnagar_skb@yahoo.de

## ABSTRACT

This paper presents the RF energy harvesting module (RECTENNA). The working range of this module includes multiple bands i.e. GSM, ISM, WLAN, and UWB band. To enhance the capturing RF power capability an array arrangement of coplanar monopole antenna has been proposed. Wilkinson power combiner has also been implemented to combine the powers of this antenna array. The RF-DC converter circuit having seven stages has also been integrated with this structure. This module produces the DC voltage of 1.8 V with respect to + 40 dB RF input. It is the unique module because it has no need of port connectors. The impedance matching of antenna and converter has been fulfilled by incorporating the passive component at the combiner's branch. The value of this passive component is kept equal to the existing value of impedance at input port of converter circuit.

## I.    INTRODUCTION

The primary focus of the Rectenna circuit is to design a suitable RF receiving antenna. For this purpose, an antenna should be designed to resonate on single, dual and multiband frequencies according to the need of the applications. The main aim of proposed work is to design a RF energy harvesting module for a wide range of frequencies which covers most of functional bands, e.g. GSM, Radio, ISM and UWB. It is well known that the RF spectrum widely employed for 900 MHz – 2 MHz (television and radio applications, GSM), 2.1GHz - 2.6 GHz (ISM band for various applications), and 3.1GHz -10.6 GHz (ultra wideband for satellite applications) [1]. Narrow band systems such as WLAN (3.1GHz - 4.4GHz), HIPERLAN (5.1GHz - 5.3GHz), C-BAND (4.4GHz - 5GHz) are also the useful bands. Keeping this in mind a compact coplanar monopole antenna has been designed and its 2 X 2 arrangement was reported [2]. For preparing the array arrangements of the antenna the literature [3-5] were profoundly studied.

This antenna works for the wide range of frequencies of the above specified bands. As reported in the structure [2] the Wilkinson power combiners were preferred because it accomplishes two essential tasks of the structure. Firstly it combines the powers of two or four antennas serially and increases the input power of RF-DC converter circuit of energy harvesting module. Secondly it considered as the element of afford, for the impedance matching of antenna port and converter circuit. The maximum power transfer concept has been chosen for this impedance matching.

The multi band operation of the RF energy harvesting module developed on a single substrate is the innovation of this paper.

## II.     ANTENNA DESIGN

In RF energy harvesting module antenna is the major functionary part. It receives the power from the electromagnetic spectrum. The received RF power is converted into the DC output voltage from the module. So the antenna should be capable enough to receive the maximum energy/power from the spectrum. As the working range of our design is preferred from 900 MHz -10 GHz, the modification in the monopole antenna structure for multi frequencies resonance has investigated. For increasing the area of receiving as well as gain of the structure the 2 x 2 array of the coplanar monopole antenna has been designed and presented [2]. In this paper [2] only the simulated results were presented. The simulated antenna structure was as shown in figure 1[2]. The power of the individual structure has been combined by the Wilkinson power combiner. The reason for selecting this combiner is that we may externally adjoin the surface mount components at the combining branch of it for impedance matching purpose.

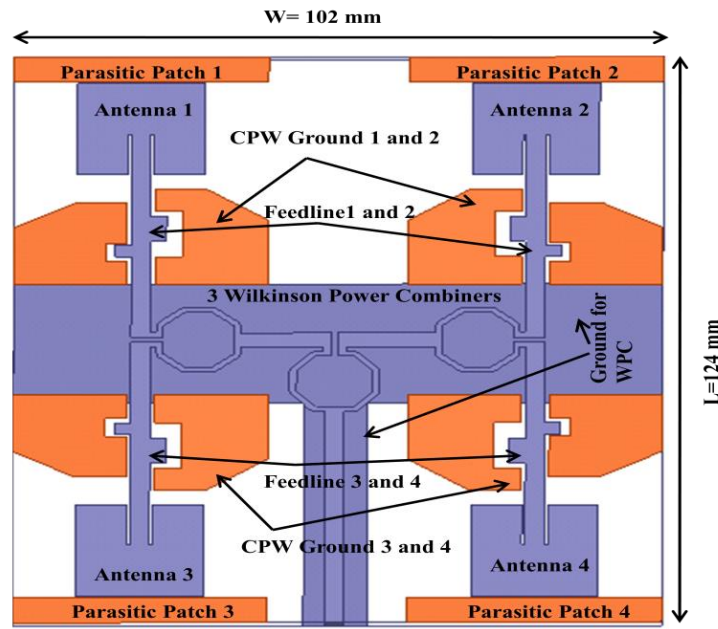

Figure 1. Simulated structure of 2 x 2 array arrangement of coplanar monopole antenna [2].

The simulated structure from licensed software HFSS version16 has been exposed resonant on the multi frequencies of different allocated bands as 900 MHz and 1.75 GHz (GSM band) 2.7 GHz (ISM), 3.5 GHz, 4 GHz and 5.2 GHz (WLAN), 6.5 GHz, 7.4 GHz, 8.5 GHz and 10 GHz (UWB).

Further in this paper all measured results have been reported for the validation of proposed antenna structure. For the measurement purpose proposed structure has been fabricated as shown in figure 2. The measured result of return loss from VNA (Keysight Technologies) for fabricated structure is shown in figure 3. The resonant frequencies of the fabricated structure has exposed on 900 MHz (GSM), 1.29 GHz (ISM), 4.1 GHz and 5.6 GHz (WLAN), 6.8 GHz and 9 GHz (UWB), that is the almost same as simulated result of the structure reported earlier [2].
.

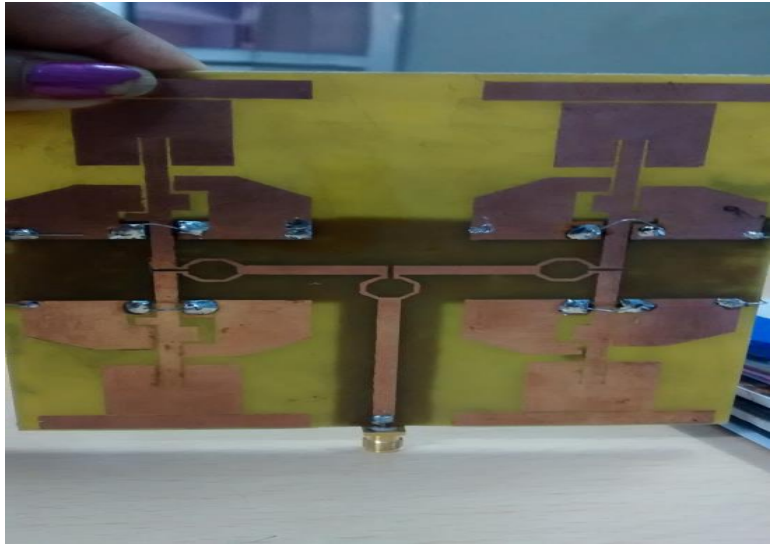

Figure 2. Fabricated structure of the 2 x 2 array of coplanar monopole antenna.

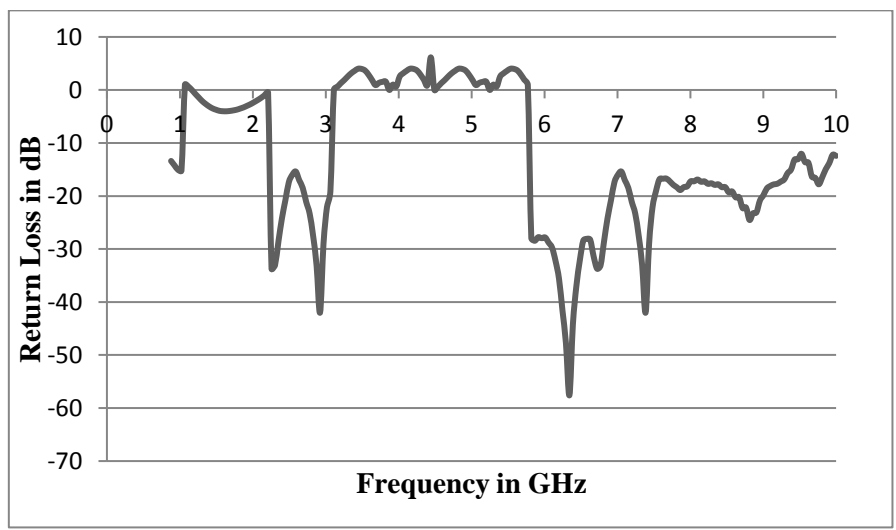

Figure 3. Measured return loss ($S_{11}$) of the 2 x 2 array of coplanar monopole antenna.

The measured H- field patterns for these frequencies are shown in figure 4(a) and E-field patterns are shown in figure 4 (b). These patterns show that the antenna supposed to be worked as omnidirectional.

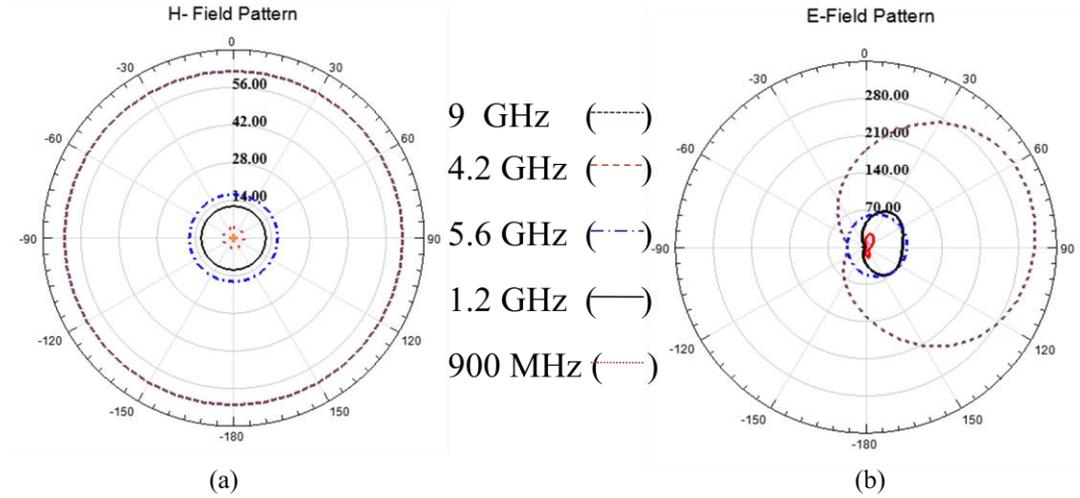

Figure 4. (a) H- field pattern (b) E- field Pattern

### III. RF-DC CONVERTER

The RF-DC converter circuit is also the important part of energy harvesting module. The output of harvester circuit is the DC voltage. The input of this module is the magnitude of AC voltage at the antenna port. This AC voltage is proportional to the captured power on the antenna from the RF spectrum. In this paper for energy conversion purpose a voltage doubler circuit has been used. The number of stages of the converter circuit may be increased for escalating the DC output voltage value. The relation of number of stages (n) and output voltage ($V_{o/p\ volt}$) of converter circuit is considered as

$$V_{o/p\ volt} = \frac{nV_{o/p\ volt}}{nR_0 + R_L} R_L$$

In this paper conversion design contains seven numbers of stages of Villard voltage doubler circuit. The main function of this conversion circuit is to convert the RF harvested energy from the ambient into the direct current (DC) voltage. This seven stage Schottky diode voltage doubler circuit is first designed, modelled and simulated on ANSYS RF circuit designer and then fabricated and tested. After that, this circuit has integrated with the same substrate of antenna. The converter circuit in this design uses zero bias Schottky diode (SMS7621-001-SOT23) from Skyworks. The simulated design of seven stage RF-DC converter circuit has been presented for this purpose [6]. The simulated DC output voltage at the first stage and seventh stage of RF-DC converter circuit from the RF circuit simulator was as 48.2 mV and 289.6 mV respectively. The fabricated circuit of seven stage RF-DC converter and its analysis has already been presented [6]. So in this paper we implemented that designed RF-DC circuit (having seven stages) [6] with the same single substrate of designed antenna array.

## IV. RF ENERGY HARVESTING MODULE

This part of paper presented a RF energy harvesting module on a single substrate working on multi bands of the frequency range 900 MHz to 10 GHz. The DC output from the multiband harvesting circuit has been presented it produces 3.4 V DC output [8]. The motive of this part is to design whole module on a single substrate for a wide range of operation and should be able to produce the DC output voltage of the value more than 1.5 V from the ambient RF spectrum. This DC voltage may be boosted by adding the booster circuit. In addition to the structure on a single substrate of RF module, it is also proposed that any value of impedance measured at the converter circuit input impedance may be matched to antenna port by using SMD (Surface mount device) resistor components. This is possible only due to the use of Wilkinson power combiner implemented at the antenna structure. No need of matching of connectors impedances. So it may be called connector-less module. This is the uniqueness of the structure.

The novel idea of impedance matching is that; for any wilkinson power combiner a given number of output ports N, an input impedance $R_S$ and output load impedance $R_L$, and the characteristic impedance of each quarter-wave section, $Z_{\lambda/4}$, is given by

$$Z_{\lambda/4} = \sqrt{N\, R_L R_S}$$

Additionally

$$R = R_L$$

the resistor (R) that connect between the common junction and the output ports are simply equal to the load impedances ($R_L$) at each output. Now considering the input impedance of RF-DC converter circuit as the value of load resistance ($R_L$) of the output of the Wilkinson power combiner. This impedance has been measured by vector network analyzer. The SMD resistor of this measured value (R) has been mounted in between the combining leg of Wilkinson power combiner. In this way we can easily match the impedance of antenna and converter circuit. This satisfies the condition of maximum power transfer.

The fabricated RF energy module on a single substrate with impedance matching arrangement is shown in figure 5. The seven stage RF-DC converter has been implemented by the combination of seven Schottky diodes and seven capacitors. The load resistance and source resistance has also been added into the circuit. The overall size of the module is 124 x 102 mm$^2$.

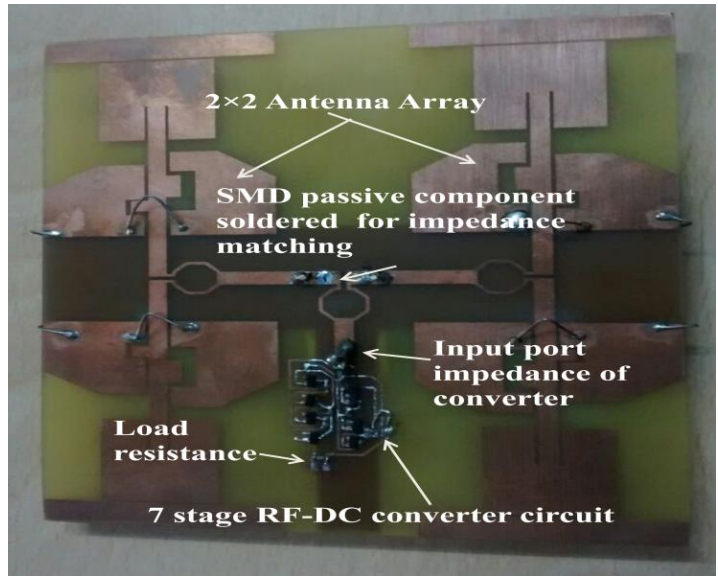

Figure 5. Fabricated RF energy harvesting module integrated on a single substrate.

The measurement with module in RF lab has been shown in figure 6. The measurement has taken at ICRS (International center for Radio Science) lab, Jodhpur. Rajasthan, INDIA. The RF source is of 40 GHz. First of all with respect to the source on the resonant frequencies of the module; power received at the antenna port has been measured on the power-meter.

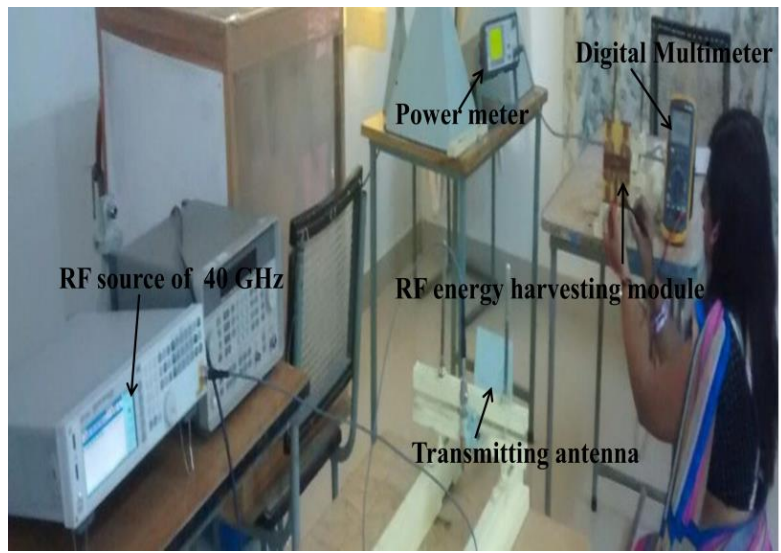

Figure 6. Measurement of output DC voltage from the module with respect to RF input voltage from RF source.

The DC output voltage at the module output with respect to RF input power varying from - 40 dB m to + 40 dB m has been plotted for five resonant frequencies (900 MHz, 1.2 GHz, 4.1 GHz,

5.6 GHz and 9 GHz). The plot of the data is as shown in figure 7. The maximum DC output is 1.823 V.

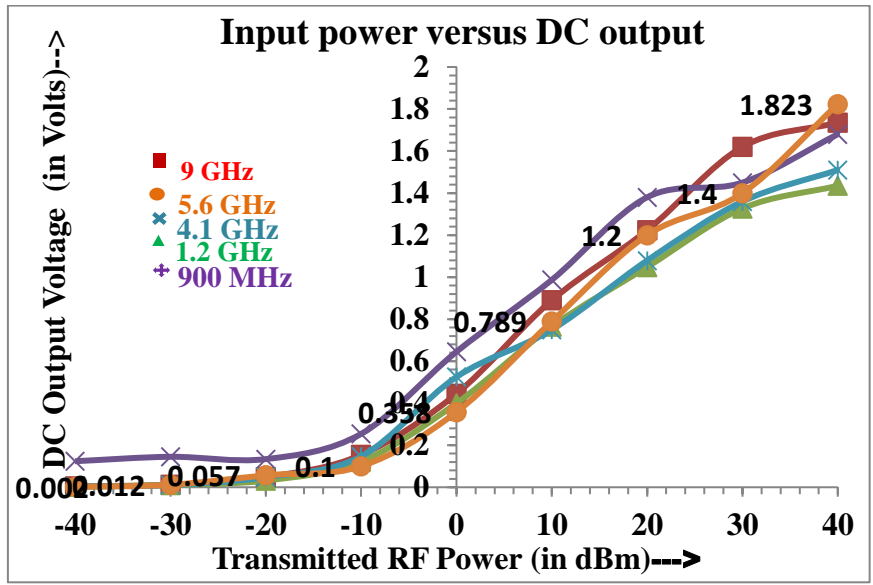

Figure 7. The transmitted RF input power versus DC output at the module graph.

The conversion efficiency (in %) of the module is calculated by the formula

$$\eta = \frac{V_{DC\ output}}{V_{input}} \times 100$$

The $V_{DC\ output}$ is the output measured by the digital multimeter and $V_{input}$ is the magnitude of the input voltage at RF power source. The efficiency for the 9 GHz frequency is shown in the figure 8. As shown in figure the efficiency increases as the RF power increases.

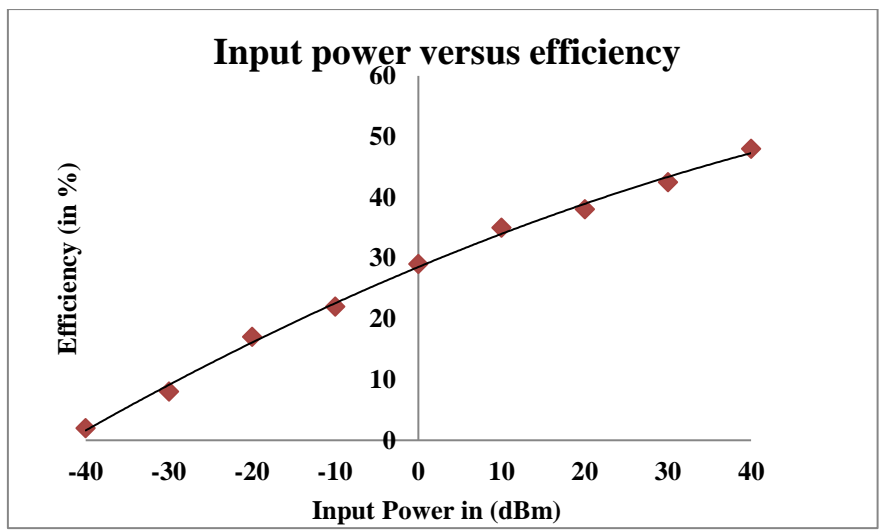

Figure 8. Input RF power versus efficiency graph.

The efficiency increases by the change in load resistance. For the input RF power +10 dBm the change in efficiency with respect to the change in load resistance is shown in figure 9. The maximum DC output voltage (1.823 V) for +40 dBm has been noted on the load resistance 22 k-ohms.

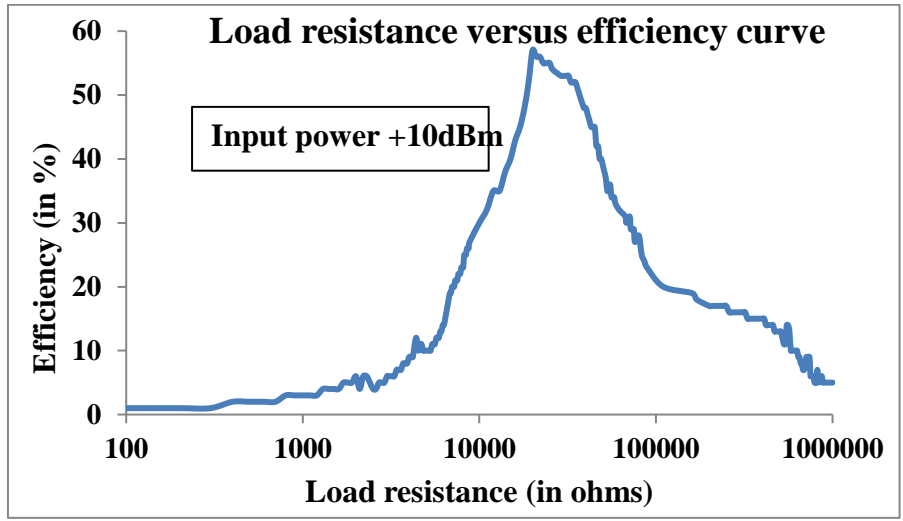

Figure 9. The load resistance versus efficiency curve.

## V. CONCLUSION

A novel RF energy harvesting module has been presented in this paper. The antenna for this module has the structure of 2× 2 array of the coplanar monopole antenna. The novelty which is proposed lies in the fact that a seven stage RF-DC converter circuit has been integrated on the same substrate of antenna. This makes the module connector-less. Another uniqueness of this module is that the impedance between the antenna and the converter circuit is matched by using the maximum power transfer concept. The Wilkinson power combiners have been specifically selected for this purpose as they combine the powers of antenna arrays. For impedance matching between the antenna output port and RF-DC converter port, the existing input impedance at the input port of the converter circuit is measured. And the component of that measured value has been mounted at the Wilkinson combining branch. This arrangement reported the matching between the antenna and circuit ports. The DC output of the module has been measured. The maximum DC output is 1.823 V with respect to +40 dB m input RF power, with the load resistance 22 k ohm for the frequency 9 GHz. This voltage may be boosted by integrating a booster circuit. The voltage may be increased by using the super capacitor at the output of module.

## VI. ACKNOWLEGEMENT

also thankful to Dr. O. P. N Calla, Director- ICRS (International Center for Radio Science) lab, Jodhpur, Rajasthan, India for providing the measurement facilities.

The author(s) declare(s) that there is no conflict of interest regarding the publication of this paper.## REFERENCES

[1] Federal Communications commission, First report and Order, (Revision of part 15 of commission's rule regarding UWB transmission system FCC, Washington, DC, pp. 02-48, 2002.

[2] Monika Mathur, Ankit Agrawal, Ghanhyam Singh, S. K. Bhatnagar, " The array structure of 2 x 2 coplanar monopole antenna with wilkinson power combiner for RF energy harvesting applications", Proc. of IEEE Xplore 2016, pp. 1-4, ISBN 978-1-5090-2807-8, 2016.
DOI: 10.1109/ICRAI.2016.7939563

[3] U. Olgun, C. C. Chen, J. L. Volakis, " Invesigation of rectenna array configurations for enhanced RF power harvesting" IEEE Antenna and Wirel. Prop. Letters, vol. 10, pp. 262-265, 2011.

[4] N. hasan and S. K. giri, "Design of low power RF to DC generator for energy harvesting applications", Int. Journal of Applied Scien. and Engg. Research, vol.1, Issue 4, pp. 562-569, 2012.

[5] T. Sakamoto, Y. Ushijima, E. Nishiyama, M. Aikawa and I. Toyoda, " 5.8 GHz series/parallel connected rectenna array using expandable differential rectenna units", IEEE Trans. on Antenna and Prop., vol. 61, no. 9, 2013.

[6] Monika Mathur, Ankit Agrawal, Ghanshyam Singh, S. K. Bhatnagar, "A Novel Design Module of RF Energy harvesting for Powering the Low Power Electronics Devices", ACM digital library- Proc. of AICTC, article no. 68, pp 1-4, 2016.
DOI: 10.1145/2979779.2979847.